\documentclass[12pt,aj]{emulateapj}

\shortauthors{Reiners \& Giampapa}

\begin{document}


\title{The Origin of Enhanced Activity in the Suns of M67}


\author{A. Reiners\altaffilmark{1}}
\affil{Institut f\"ur Astrophysik, Georg-August-Universit\"at,
 Friedrich-Hund-Platz 1, 37077 G\"ottingen, Germany}
\email{Ansgar.Reiners@phys.uni-goettingen.de}
\and
\author{M.S. Giampapa}
\affil{National Solar Observatory\altaffilmark{2}, National Optical Astronomy
 Observatory\altaffilmark{2}, 950 North Cherry Avenue, P.O. Box 26732, 
Tucson, AZ 85726-6732}
\email{giampapa@noao.edu}


\altaffiltext{1}{Emmy Noether Fellow}
\altaffiltext{2}{The National Solar Observatory and the National Optical 
Astronomy Observatory are each operated by the
Association of Universities for Research in Astronomy, Inc. (AURA) under
respective cooperative agreements with the National Science Foundation}


\begin{abstract}
  We report the results of the analysis of high resolution
  photospheric line spectra obtained with the UVES instrument on the
  VLT for a sample of 15 solar-type stars selected from a recent
  survey of the distribution of H and K chromospheric line strengths
  in the solar-age open cluster M67.  We find upper limits to the
  projected rotation velocities that are consistent with solar-like
  rotation (i.e., $v\,\sin{i} \la$ 2--3\,km\,s$^{-1}$) for objects
  with Ca II chromospheric activity within the range of the
  contemporary solar cycle. Two solar-type stars in our sample exhibit
  chromospheric emission well in excess of even solar maximum values.
  In one case, Sanders~1452, we measure a minimum rotational velocity
  of $v\,\sin{i} = 4 \pm 0.5$\,km\,s$^{-1}$, or over twice the solar
  equatorial rotational velocity. The other star with enhanced
  activity, Sanders~747, is a spectroscopic binary.  We conclude that
  high activity in solar-type stars in M67 that exceeds solar levels
  is likely due to more rapid rotation rather than an excursion in
  solar-like activity cycles to unusually high levels.  We estimate
  an upper limit of 0.2\,\% for the range of brightness changes
  occurring as a result of chromospheric activity in solar-type stars
  and, by inference, in the Sun itself. We discuss possible
  implications for our understanding of angular momentum evolution in
  solar-type stars, and we tentatively attribute the rapid rotation in
  Sanders~1452 to a reduced braking efficiency.
\end{abstract}





\section{Introduction}

In a survey of the Ca II H and K core strengths of a sample of 60
solar-type stars in the solar-age, solar-metallicity open cluster M67,
\citet{Giampapa06} found that the distribution of the HK index -- a
measure of the strength of the chromospheric H and K cores -- is
broader than the distribution seen in the contemporary solar cycle.
Significant overlap between the HK distribution of the solar cycle and
that for the sun-like stars in M67 is seen with over 70\% of the solar
analogs exhibiting Ca II H+K strengths within the range of the modern
solar cycle. About $\sim$ 10\% are characterized by high activity in
excess of solar maximum values while approximately 17\% have values of
the HK index less than solar minimum.

In view of these results, a key question that arises is whether the
distribution of the HK index in M67, and especially the occurrence of
solar-type stars in this cluster with levels of activity exceeding
solar maximum, is due to excursions in the amplitude of otherwise
solar-like cycles or whether the high HK values in some solar-type
members are simply the result of their more rapid rotation. The former
possibility is reminiscent of the prolonged episode of quiescence
known as the Maunder Minimum where sunspots vanished, coinciding with
the Little Ice Age \citep{Foukal90}, as well as the extended period of
high solar activity, as inferred from isotopes in the terrestrial
record, known as the Medieval Solar Maximum during the so-called
Medieval Warm Epoch \citep{Jirikowic94}. In the case of the latter
possibility of more rapid rotation, we know that atmospheric activity
associated with surface magnetic fields, in general, increases with
increasing equatorial rotation velocity in late-type stars (e.g.,
Skumanich 1972; Pallavicini et al. 1981; Baliunas et al. 1995;
Pizzolato et al. 2003).

\begin{figure*}
\centering
\includegraphics[width=.9\textwidth]{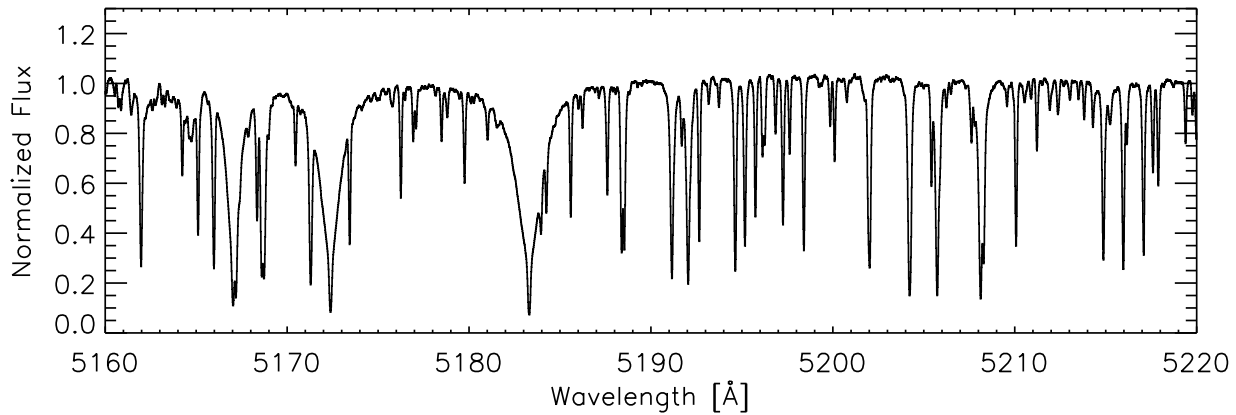}
\caption{\label{fig:Spectrum}Spectral window of our Ganymede spectrum.}
\end{figure*}

The presence of active solar-type stars in the solar-age cluster M67
is particularly surprising. It is therefore crucial to determine
whether it is rapid rotation that is the origin of the relatively
enhanced activity, or if the rotation of these objects is solar-like
but their cycle variations extend to relatively higher amplitudes than
seen in the Sun. If it is more rapid rotation that is the natural
origin of the higher Ca II core emission then this result would invite
further investigation of the angular momentum evolution of M67 in
contrast to that of the more slowly rotating Sun and other solar-age G
dwarfs. However, if rotation velocities are solar-like at
$\sim$2\,km\,s$^{-1}$ then this would suggest that excursions in the
cycle variation of solar-type stars that are significantly in excess
of contemporary solar maximum values can occur in Sun-like stars and,
by implication, in the Sun itself. We therefore discuss in this
investigation the results of our attempt to measure projected rotation
velocities in selected M67 solar-type stars from the sample of
\citet{Giampapa06}. The sample selection and the acquisition of the
observations are discussed in $\S\S$2-3 while the novel approach to
the analysis is given in $\S$4. The results are presented in $\S5$ and
a discussion followed by our conclusions are given in $\S\S$6 and 7,
respectively.

\section{Sample selection}

We selected a sample of 15 sun-like stars from \citet{Giampapa06}. Our
targets cover a range in $B-V$ of 0.62--0.67, which corresponds to
spectral types G2--G6 or effective temperatures of 5700--5850\,K
\citep{Kenyon95}. \citet{Giampapa06} found that the majority of stars
in M67 exhibits Ca\,H\&K emission similar to the Sun; most of their
targets populate the range 155--260\,m\AA\ in HK index.  As defined in
\citet{Giampapa06}, this index is is the sum of the relative
intensities in 1 {\AA} bandpasses centered at the H and K lines in the
calibrated spectrum.  Two outliers with exceptionally strong Ca II
emission were found, Sanders~747 (354\,m\AA) and Sanders~1452
(414\,m\AA). Our sample consists of 13 stars with HK indices similar
to the Sun plus the two stars with higher HK indices. Additionally, we
obtained a spectrum of the Sun through the same instrumental setup by
observing its light reflected from Jupiter's moon Ganymede.

\section{Data}

The data were obtained using ESO's high resolution spectrograph UVES
at the VLT in its red arm centered at 580\,nm. This setup covers the
wavelength range 480\,nm to 680\,nm on two chips with a gap at
575--585\,nm. To achieve the highest possible spectral resolution, we
used the image slicer \#3 providing a resolving power of $R \sim
110\,000$. After exposure times between 30\,min and 1\,h per star, the
spectra from the blue chip have signal-to-noise ratios of about 40 per
resolution element. The spectrum of Ganymede attained a SNR above 400
after 20\,s. A part of our Ganymede spectrum from the blue chip is
shown in Fig.\,\ref{fig:Spectrum}.

The part of the spectrum obtained on the red chip unfortunately is
compromised by an interference pattern which is most likely related to
electronic noise. Such an interference pattern with a peak-to-peak
amplitude of about 1~ADU was present from the beginning of UVES
operations\footnote{We thank the ESO user support department for their
help with the data. A replacement of the red chip of UVES is
foreseen in May 2009.}. In our case, the reason for the noise
pattern seems to be related to the interference pattern, but in our
data the noise has grown much larger. This pattern is difficult to
correct for, and we decided not to use data from the red chip for our
analysis; the available spectral range on the blue chip carries
sufficient information and the red part is not crucial for our
purpose.

\section{Analysis}

At a SNR of 40 and a resolving power of $R \sim 110\,000$, i.e.
$\Delta v \sim 2.7$\,km\,s$^{-1}$, the difference between sun-like
stars rotating at 2\,km\,s$^{-1}$ and 4\,km\,s$^{-1}$ cannot be
detected with high confidence. The main obstacle is the large
intrinsic width of spectral lines in sun-like stars, which due to
temperature broadening and turbulent motions have FWHM exceeding
6\,km\,s$^{-1}$ \citep[e.g.][]{Gray05}. Thus, at a sampling rate
slightly better than 3\,km\,s$^{-1}$, the difference between
additional rotational broadening of 2 and 4\,km\,s$^{-1}$ is very
subtle.

To overcome the relatively low SNR, we derive a mean line broadening
over a large spectral range including several hundred spectral lines
\citep[see also][]{Reiners03a, Reiners03b}. By performing a
least-squares deconvolution process, we search for line broadening
that is common to all lines that we take into account. In other words,
we search for the broadening function that, applied to a template
spectrum of unbroadened spectral lines, produces the best fit to our
data. As a template, we start with a ``$\delta$-template'' that is
non-zero only at the position of spectral lines and zero elsewhere
(for technical reasons, we do not fit to the normalized spectrum
$f(\lambda)$, but to $1-f(\lambda)$ instead). The intensity at the
position of spectral lines is estimated from the modelled central line
depth taken from the Vienna Atomic Line Database \cite[VALD,][]{VALD}.
Before the deconvolution process, we apply a ``thermal'' broadening to
each line in the template, i.e., we convolve each line with a Gaussian
according to the expected motion of the gas at given atomic mass and
gas temperature. This broadening is different between lines of
different atoms and hence has to be included before the deconvolution.
With this template in hand, we search for the function that, convolved
with our template, provides the best fit to the data. In this
function, each of the typically 60--80 pixels is a free parameter.

It is important to realize that the described analysis is mainly a
technical procedure. We do not model the full set of parameters
relevant for spectral line broadening, which include for example
turbulence velocities, temperature structure and density
stratification. Instead, we assume that our sample stars have
atmospheres that are approximately the same at least on the level
interesting for this analysis, and that the main difference in
spectral line width is due to rotation.

We utilize in our analysis the spectral range 5400--5750\,\AA, where
more than 300 spectral lines in this region are included in the fit.
In order to accurately recover the shape of the broadening function,
it is important that the equivalent widths of the spectral lines are
chosen correctly. Equivalent widths are estimated from VALD data in
the first step, but this usually does not provide an adequate set of
lines. Moreover, at this level of accuracy, stars are very different
in the distribution of line depths. We adjust for the distribution of
line depths by an iterative procedure fitting line depths and
broadening function alternately. This means that after the initial
deconvolution described above, we run the same fitting procedure but
now we fit for all line depths instead of the broadening function,
which we keep fixed. With the new set of line depths we repeat the
fitting of the broadening function and so forth until a stable
solution is reached.

\subsection{The broadening function of the Sun}

\begin{figure}
\plotone{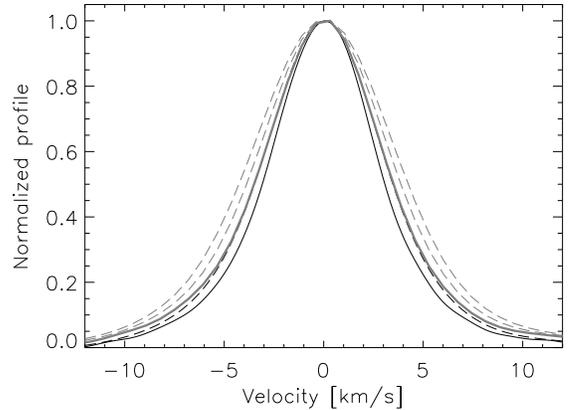}
\caption{\label{fig:Sun}Broadening function of the Sun from the
solar flux atlas (thin solid line). Thick grey line: Ganymede
broadening function from UVES observations. Thin black dashed
line: Broadening function from the solar flux atlas (thin solid line)
artificially broadened by a Gaussian instrumental profile with
R=110\,000. Dashed grey lines: Ganymede broadening function
artificially broadened to show a Sun rotating at 3 and 4 km/s.}
\end{figure}

In order to test our deconvolution procedure on real data, we applied
the very same process to the solar flux atlas provided by NSO/NOAO
\citep[$R \sim 400,000$, ][]{KPA}, and we observed the Sun with UVES
using the same instrument setup as for our M67 targets. This was done
because the effect of our deconvolution procedure on spectra from very
slow rotators is not straightforward to estimate. In particular, this
test reveals whether any spurious broadening is introduced by the
deconvolution process. Furthermore, we use this test to investigate
the influence of the UVES image slicer on the line broadening. The
results of the deconvolution procedure in the two solar spectra are
shown in Fig.\,\ref{fig:Sun}.

First, one can see that the deconvolution process produces a smooth
broadening function with a FWHM of roughly 6\,km\,s$^{-1}$ from the
solar flux atlas (thin solid line). The wings of this function are
relatively broad extending out to $\pm 10$\,km\,s$^{-1}$. We compared
this profile to a few absorption lines in the spectrum finding that
the line width and shape of the broadening profile does not
significantly differ from those of individual lines. At a rotation
velocity of $v\,\sin{i} = 1.8$\,km\,s$^{-1}$, the main effect of line
broadening comes from turbulent gas motion. This effect is well known
and should not be further studied here, but it is important to realize
that even at very high spectral resolution, the slow rotation of the
Sun is barely detectable in its spectrum.

The broadening profile of the Sun seen through the UVES instrument is
also shown in Fig.\,\ref{fig:Sun} (thick grey line). As expected, it
is somewhat broader than the one from the solar flux atlas. We
artificially broadened the broadening function from the solar flux
atlas to the spectral resolving power of UVES ($R \sim 110\,000$). The
result is overplotted as a black dashed line in Fig.\,\ref{fig:Sun}.
This line resembles remarkably well the broadening profile derived
from our UVES observations (the small difference beyond $\pm
6$\,km\,s$^{-1}$ is probably an effect of the image slicer). Thus, we
are confident that the UVES instrument does not introduce additional
broadening on a scale that is relevant for our investigation. Note
that the difference between the broadening profiles introduced by
spectral resolutions that differ by about a factor of ten ($\sim 10^6$
for the solar flux atlas and $\sim 10^5$ for UVES) is relatively
little. This is another consequence of the strong broadening due to
turbulence, which at a resolution of $R \sim 110\,000$ is already
sampled quite well.

\subsection{The effect of rotation}

To display the effect of rotation on the broadening functions in
sun-like stars, we artificially broadened the spectrum of Ganymede to
construct the broadening function of a sun-like star rotating at
projected velocities of $v\,\sin{i} = 3$ and $4$\,km\,s$^{-1}$.
Artificial broadening is done by a convolution of the spectrum with a
broadening function $G(v)$\citep[e.g., Eq. (18.14) in ][]{Gray05}. We
took into account the fact that the Sun is rotating ($v\,\sin{i} \sim
1.8$\,km\,s$^{-1}$), i.e., we approximate rotational broadening as a
net effect of two consecutive broadening steps. The total broadening
of $v\,\sin{i}$ given by the quadratic sum of two individual steps. In
our case, the two steps are (1) the broadening introduced by solar
rotation, and (2) a ``reduced'' artificial broadening. In other words,
we use $v\,\sin{i}_{\rm reduced} = \sqrt{(v\,\sin{i})^2 - (1.8\,{\rm
    km\,s^{-1}})^2}$ = 2.4 and 3.6\,km\,s$^{-1}$ for the cases of
$v\,\sin{i}$ = 3 and 4\,km\,s$^{-1}$, respectively. We note that
quadratically adding rotation velocities is only an approximation; the
result of two artificial broadening processes cannot be described by
one ($G(v_1) \ast G(v_2) \neq G(v_3)$). However, we are mainly
interested in the differential comparison between the profiles. The
correction we apply to the absolute rotational velocity is on the
order of $10\,\%$, and we estimate that the \emph{differential} error
introduced by the imprecise treatment of consecutive rotations is on
the order of $1$\,\% or less.

The two cases of $v\,\sin{i}$ = 3 and 4\,km\,s$^{-1}$ are shown in
Fig.\,\ref{fig:Sun} as grey dashed lines. They are clearly
distinguishable, which demonstrates that it is possible to
differentiate between rotational broadening of 2, 3, and 4
km\,s$^{-1}$ in our data.

\section{Results}

As explained in the former sections, we derived the broadening
functions of all 15 members of M67 that we observed along with that
for the Sun.  In all cases, the iteration process converged to a
stable solution providing a smooth broadening function. The stars of
our sample and the solutions from our analysis are summarized in
Table\,\ref{tab:results}. Two stars of our sample show exceptionally
high activity, i.e., a Ca\,HK index higher than 300\,m\AA. The
principal result of our analysis is that these two stars also show
peculiar broadening functions. All other stars exhibit activity
indices consistent with solar values observed during the solar
activity cycle. These stars have broadening functions resembling the
solar one implying they are consistent with very slow rotation, i.e.,
$v\,\sin{i} \la 2$\,km\,s$^{-1}$.

\begin{deluxetable}{rccc}
\tablecaption{\label{tab:results}Results}
\tablewidth{0pt}
\tablehead{\colhead{Sanders} & & \colhead{HK} & \colhead{$v\sin{i}$}\\
Number & $(B-V)_{\rm 0}$ & [m\AA] & [km/s]}
\startdata
746 & 0.66 & 209 & -- \\
747 & 0.65 & 354 & SB2 \\
770 & 0.63 & 195 & -- \\
777 & 0.63 & 208 & -- \\
785 & 0.65 & 221 & -- \\
802 & 0.67 & 193 & -- \\
991 & 0.63 & 179 & -- \\
1048 & 0.64 & 188 & -- \\
1106 & 0.67 & 168 & -- \\
1203 & 0.67 & 218 & -- \\
1218 & 0.63 & 194 & -- \\
1246 & 0.64 & 187 & -- \\
1452 & 0.62 & 414 & $4\pm0.5$ \\
1462 & 0.63 & 193 & -- \\
1477 & 0.67 & 181 & -- \\
\enddata
\end{deluxetable}

\subsection{The spectroscopic binary Sanders~747}

\begin{figure}
\plotone{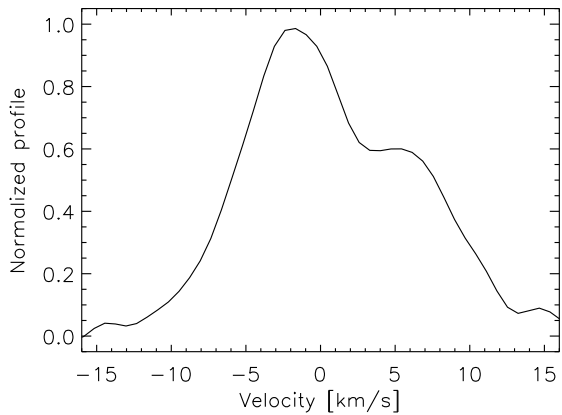}
\caption{\label{fig:S747}Broadening function of the spectroscopic
binary Sanders~747.}
\end{figure}

The first star that shows a peculiar broadening function is
Sanders~747. We show the broadening function in Fig.\,\ref{fig:S747}.
It clearly resembles the shape of a spectroscopic binary profile.
Because both components are seen in the deconvolved broadening
function, their spectral types cannot be too different (otherwise one
of the components would not be visible because it would not display lines
at the positions of the template). The components have different
maximum intensities, which could either mean that one component is
brighter (or has more lines matching the template), or that one
component is more rapidly rotating causing shallower lines. In fact,
the slope of the stronger component (left wing of the broadening
function) is steeper than the slope of the weaker component (right
wing). This probably indicates a higher rotation velocity in one of
the components. This high rotation velocity could be the reason for
the enhanced activity observed in Sanders~747, but interactions
between the two binary components may also be responsible through
other mechanisms.

\subsection{Rapid rotation in Sanders~1452}

\begin{figure}
\plotone{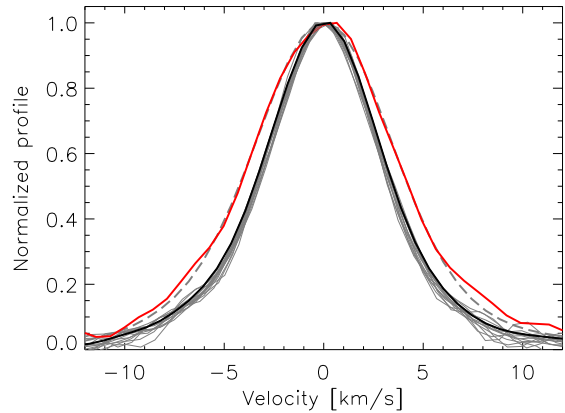}
\caption{\label{fig:Profiles}Broadening profiles of M67 stars (thin
grey lines and thick red line) and of the Sun through Ganymede
(thick black line). Dashed grey line: Ganymede spectrum
artificially broadened to reproduce a $v\,\sin{i} =
4$\,km\,s$^{-1}$ star (the same as the widest line in
Fig.\,\ref{fig:Sun}).}
\end{figure}

The broadening functions of all single stars of our sample are shown
in Fig.\,\ref{fig:Profiles}. Thin grey lines show those of stars that 
are consistent with very slow rotation ($v\,\sin{i} \la
2$\,km\,s$^{-1}$). It is remarkable that all 13 profiles match each
other very closely. They also closely resemble the broadening function
of the Sun (Ganymede, thick black line).

The only star exhibiting a broadening function that is significantly
wider than the others is Sanders~1452 (red line). We have overplotted
an artificially broadened profile resembling a star rotating at
$v\,\sin{i} = 4$\,km\,s$^{-1}$ (grey dashed line, which is identical
to the widest profile shown in Fig.\,\ref{fig:Sun}). The profile
expected from a star rotating at $v\,\sin{i} = 4$\,km\,s$^{-1}$ is a
very good match to the profile of Sanders~1452, with only the far
wings beyond $\pm 6$\,km\,s$^{-1}$ differing slightly. This again may
be an effect of the image slicer, but the far wings of the deconvolved
broadening function are also less well defined due to the existence of
uncaptured line blends and limited SNR in the data.

From the close match between the broadening function of Sanders~1452
and the artificially broadened versions of the Ganymede spectrum, we
derive a value for the projected rotation velocity of Sanders~1452 of
$v\,\sin{i} = 4 \pm 0.5$\,km\,s$^{-1}$. The uncertainty is an
empirical estimate deduced from a comparison with broadening functions
with different values of $v\,\sin{i}$. Broadening functions with
$v\,\sin{i} = 3.5$ and 4.5\,km\,s$^{-1}$ can be distinguished from the
data.

\section{Discussion}

The results of our detailed spectroscopic analysis enable us to
discuss the range of chromospheric activity seen in M67 in conjunction
with the additional information on rotation, at least for a
representative subset of solar-type stars in M67. We therefore
briefly discuss in the following some relevant issues in this context,
namely, (1) the potential range of brightness variability in sun-like
stars at solar age and, by inference, in the Sun itself, and (2) some
facets of angular momentum evolution in solar-type stars, including
the possible origin of relatively rapid rotation at the age of the M67
open cluster.

\subsection{The Range of Activity and Brightness Variability in
  Sun-like Stars}

The potential excursions of the activity cycles of the M67 solar-like
stars and possibly the Sun itself to exceptionally high values, as
inferred from the HK distribution of the M67 solar-type stars given in
\citet[][their Fig. 3]{Giampapa06}, now must be considered in the
light of the rotation measures and estimates given herein. In
particular, since S1452 with a mean HK index of 414 m{\AA} is rotating
at more than twice the equatorial solar rotation velocity, and S747
with a mean HK = 354 m{\AA} is a spectroscopic binary, the implication
is that excursions in the cycles of M67 solar-type stars and the Sun
itself appear to be less than about HK $\sim$ 250 m{\AA}, i.e., the
next highest HK index found in the Giampapa et al. sample, which is
roughly 10\% higher than the representative maximum value seen in the
modern solar cycle of HK $\approx$ 225 m{\AA}. Within the reported
$B-V$ color range of the Sun of about 0.63 to 0.67
\citep{VandenBerg84}, the only star in the Giampapa et al. sample that
exceeds the maximum solar HK index (S 1014) also is a short-period
binary with a period of 16.2 days and mean HK = 248 m{\AA}.  The
determinination of a definitive upper-limit to the HK index for single
sun-like stars at solar age and metallicity will require a more
extensive survey of rotation in the M67 solar-type stars.

Given our results and the well-known correlation between variations in
chromospheric emission and changes in the solar irradiance or,
correspondingly, in stellar brightness, it is of interest to consider
the implications of the plausible upper limit to chromospheric
emission in solar-type and solar-age stars, as inferred from the M67
sample, for the possible range of brightness variability that may
occur. Using the color-dependent calibration approach described in
\citet{Giampapa06} and adopting a color of $B-V = 0.65$ as
representative of analogs of the Sun, we find that the upper limit
given above of HK $\simeq$ 250 m{\AA} corresponds to log
R$^{\prime}_{HK} \sim -4.70$, where R$^{\prime}_{HK}$ is the ratio of
the total chromospheric H and K emission core flux to the stellar
bolometric flux, corrected for the non-chromospheric (photospheric)
contribution. Inspection of the results of long-term, high precision
photometry of solar-type stars by \citet[][their Fig. 17]{Lockwood97}
as a function of the parameter R$^{\prime}_{HK}$ suggests that this
level of activity would correspond to an annual mean level of rms
brightness variations of roughly 0.002 mag, i.e. 0.2\% variability in
the brightness as recorded in the Str\"omgren $b$ and $y$ bands , or
about twice the $\sim 0.1\%$ variation in total irradiance that has
been measured thus far for the contemporary Sun during the solar
cycle. Therefore, we suggest that $\sim 0.2\%$ represents an upper
limit to the likely excursion of the solar luminous output as a result
of enhanced levels of magnetic activity.

We can refine this estimate of the upper limit further by noting that
the brightness changes given by \citet{Lockwood97} were for the mean
variation of the sum of the Str\"omgren $b$ and $y$ bands. Hence, the
variation of the total irradiance must be less than what is observed
in these visible spectral bands. \citet{Radick98} estimated (for small
variations) a factor for converting between a fractional change in
bolometric flux into the corresponding magnitude difference in $(b +
y)/2$. Adopting their conversion factor of 1.39 and the estimate of
brightness variations of 0.002\,mag given above yields an estimate of
0.14\% for the upper limit for variations in the bolometric flux.
This is only slightly larger than the mean variation of 0.1\% in the
total solar irradiance observed during the course of the solar cycle.

\subsection{Angular Momentum Evolution in M67}

The relatively more rapid rotation of S1452 invites further
consideration in the context of angular momentum evolution and the
determination of stellar ages based on rotation, known as
"gyrochronology" \citep{Barnes07}. At its measured (projected)
rotation velocity and assuming a stellar radius close to solar, the
``gyro-age'' of Sanders~1452 is $1.0 \pm 0.2$\,Gyr \citep{Barnes07}.
This value is an upper limit because we measure only the projected
velocity $v\,\sin{i}$. A more direct comparison can be obtained with
the $v\,\sin{i}$--age correlation given by \citet[][their Fig.
9]{Pace04}. The inferred age (upper limit) of S1452 based on the three
possible power law fits adopted by Pace \& Pasquini is in the range of
1.2--1.5 Gyr. In either approach, the rotation-based age estimate for
S1452 is in vivid contrast to the age range for M67 of 3.5--4.8 Gyr
\citep{Yadav08} or that of the Sun, namely, 4.57\,Gyr
\citep[e.g.,][]{Bonanno02, Baker05} and the similar solar age implied
for the slow rotators of our sample.

Given its higher rotational velocity, S1452 therefore represents an
alternative path for angular momentum evolution among single stars in
this solar-age cluster. In view of the relevance to the calibration of
age--rotation laws and even the applicability of age--rotation
relations to individual objects, a discussion of our results in the
context of rotational evolution is merited.

In brief overview, current models of angular momentum evolution
\citep[e.g.][]{Bouvier97, Allain98} suggest that stars rotating more
rapidly on the ZAMS had a short-lived disk with a correspondingly
shorter disk-locking time during their pre-main sequence phase than
did their more slowly rotating counterparts. In addition, these stars
may have had a higher initial angular momentum. However, after arrival
on the ZAMS, models suggest that fast rotators spin down even more
rapidly due to a more efficient magnetized wind \citep{Kawaler88},
that, in turn, dominates the angular momentum evolution of solar-type
stars. As a consequence, solar-type stars should converge to the same
rotation velocities at the age of the Sun (or M67) independent of
their ZAMS rotation rates. This scenario would imply that magnetic
braking did not operate as efficiently in S1452 as in the other
sun-like stars in our sample.

An additional perspective is provided by \citet{Bouvier08} who
considers the effects of stellar rotational history on lithium
depletion. In particular, \citet{Bouvier08} finds that rotational
mixing is governed by the rotational shear at the base of the
convective envelope, which, in turn, depends on the degree of
core-envelope coupling. While this model is primarily applicable
during the first $\sim$ 1 Gyr in solar-type stars, the potential
extension of the hypothesis of core-envelope decoupling to main
sequence rotational evolution may be briefly considered. The result in
his models is that lithium is more severely depleted in slow rotators
on the ZAMS than fast rotators because slow rotators are characterized
by stronger core-envelope decoupling that in a non-specific way leads
to more efficient rotation-induced mixing. Conversely, fast rotators
have less rotational shear in this model with correspondingly less
efficient rotational mixing and, as a result, a lower rate of lithium
depletion. In this regard, the recent investigation of candidate
"solar twins" in M67 by \citet{Pasquini08} is relevant.

In particular, the more rapidly rotating S1452 has a higher lithium
abundance at log N(Li) = 1.0 than do 7 of the 10 best solar twin
candidates identified by \citet{Pasquini08}.  This would suggest in
the above model context of rotation-induced mixing that a spin-down
with core-envelope decoupling dominated the rotational evolution of
the solar twins. However, we note the important caveat that S1452 is
slightly warmer than the solar twins.  Therefore, it should not be
surprising on this basis alone that the lithium depletion rate would
be higher in the solar twins than in S1452.

A more appropriate comparison is with M67 stars characterized by
estimated effective temperatures that are closer to that of our
object.  In this regard, S1452 has a lower lithium abundance than most
of the stars in its bin of effective temperature between 5900\,K to
5950\,K.  Furthermore, of the seven stars in our Table 1 that are in
common with \citet{Pasquini08}, five have Li abundances greater than
S1452 and two have lower abundances that are sun-like. These objects
are slow rotators--an assertion that is supported by our measured
upper limits of $v\,\sin{i} \la 2$\,km\,s$^{-1}$ and their sun-like HK
values.  Furthermore, they have slightly lower effective temperatures
which would suggest a relatively larger lithium depletion rate than in
S1452.

Therefore, it seems unlikely in the above model context that there was
a prolonged period on the main sequence of decoupled core-envelope
spin-down for the slow rotators and, conversely, an extended period of
coupled core-envelope rotational evolution for objects such as S1452.
Moreover, the hydrodynamic or magnetohydrodynamic processes involved
in any core-envelope decoupling and their association with the mixing
of chemical species remain to be developed in the model advanced by
\citet{Bouvier08}. At this point, it would seem that the most natural
qualitative explanation of the relatively higher rotational velocity
of S1452 is that wind braking is less efficient in this object
possibly as a result of a field configuration that is dominated by
higher multipole moments, which would also account for the higher
level of chromospheric activity. But why this object should differ in
this way from the other solar-type stars in M67 in its rotational
evolution is not at all clear. We only note that S1452 is the earliest
star in the sample ($B-V = 0.62$ while all others have $B-V = 0.63 -
0.67$). Hence, its higher rotation rate could be a natural consequence
of a mass-dependence for rotational evolution on the M67 main
sequence. A more extensive $v\,\sin{i}$ survey, particularly in a
narrow color bin centered on the color of S1452, would have to be
performed to further examine this possibility.

\section{Conclusions}

We find on the basis of the analysis of high resolution spectra of a
subset of solar-type stars from the H \& K survey of these objects in
M67 by \citet{Giampapa06} that stars with levels of chromospheric
activity within the range of the contemporary cycle appear to have
rotational velocities that are solar-like. At least one solar-type
star with an HK index well in excess of the modern solar maximum is
rotating more rapidly. Specifically, we measured a value for the
projected rotation velocity of Sanders~1452 of $v\,\sin{i} = 4 \pm
0.5$\,km\,s$^{-1}$, or more than twice the solar equatorial rotation
velocity. Another object in our sample, Sanders~747, is found to be a
spectroscopic binary and is likely characterized by more rapid
rotation than the Sun in at least one of its components. In view of
these results, we conclude that the occurrence of high-activity levels
in excess of solar maximum values for solar-type stars in M67 is most
likely due to relatively faster rotation rather than to an excursion
of a solar-like cycle to high values at solar rotation rates. Based
on the correlation found between relative chromospheric emission and
mean brightness variations we estimate an upper limit of about 0.2\%
in the level of broad-band variability in solar-type stars at solar
ages and, by inference, in the Sun itself.

After a consideration of current models for angular momentum evolution
for low mass stars, we tentatively attribute the more rapid rotation
of solar-age stars such as Sanders~1452 to a reduced efficiency of
braking due to a magnetized wind. Alternatively, the rapid rotation
also could result simply from the particular mass-dependence of
rotational braking in M67, as is seen in other clusters. A more
comprehensive $v\,\sin{i}$ survey in M67 with specific emphasis on
stars slightly warmer than the Sun will have to be performed to
explore this hypothesis further.

\acknowledgements

The results presented in this investigation were based on observations 
obtained at the European Southern Observatory,
Paranal, Chile, PID 080.D-0139. A.R. has received research funding
from the DFG as an Emmy Noether fellow (RE 1664/4-1). The data utilized
herein from the FTS instrument operated at the NSO/Kitt Peak McMath-Pierce
Solar Telescope Facility were produced by NSF/NOAO. We acknowledge partial 
support of the publication of this work by a grant from the NASA/Astrobiology
Institute to the University of Arizona and the NOAO.

\end{document}